\documentclass[%showpacs, 
twocolumn, preprintnumbers, amsmath,amssymb]{revtex4}
\usepackage{graphicx}% Include figure files
\usepackage{dcolumn}% Align table columns on decimal point
\usepackage{bm}% bold math

\newcommand\ee{\end{equation}}
\newcommand\be{\begin{equation}}
\newcommand\eea{\end{eqnarray}}
\newcommand\bea{\begin{eqnarray}}
\newcommand{\sfrac}[2]{{\textstyle\frac{#1}{#2}}}

\def\d{\partial}
\def\l{\left(}
\def\r{\right)}

\newcommand{\bg}{\begin{gather}}
\newcommand{\eg}{\end{gather}}
\newcommand{\bseq}{\begin{subequations}}
\newcommand{\eseq}{\end{subequations}}

\begin{document}

%\preprint{}

\author{Sergei Dubovsky}
\email{dubovsky@nyu.edu}

\affiliation{%
Physics Department and Center for Cosmology and Particle Physics,\\
New York University, New York, NY 10003, USA
}%

\author{Lam Hui}
\email{lhui@astro.columbia.edu}

\affiliation{%
Physics Department and Institute for Strings, Cosmology, and Astroparticle Physics,\\
Columbia University, New York, NY 10027, USA
}%

\author{Alberto Nicolis}
\email{nicolis@phys.columbia.edu}

\affiliation{%
Physics Department and Institute for Strings, Cosmology, and Astroparticle Physics,\\
Columbia University, New York, NY 10027, USA
}%

\title{Effective field theory for hydrodynamics: \\ Wess-Zumino term and anomalies in  
 two spacetime dimensions}% Force line breaks with \\

\date{\today}% It is always \today, today,

\begin{abstract}
We develop the formalism that incorporates  quantum anomalies in the effective field theory 
 of non-dissipative fluids. 
 We consider the effect of adding a Wess-Zumino-like term to the low-energy effective action  
 to account for anomalies. In this paper we restrict to two spacetime dimensions.
We find modifications to the constitutive relations for the current and the stress-energy tensor, 
and, more interestingly, {\em half} a new propagating mode (one-and-a-halfth sound): a left- {\em or} right-moving wave with propagation speed that goes to zero with the anomaly coefficient. Unlike for the chiral magnetic wave in four dimensions, this mode propagates even in the absence of external fields. 
We check our results against a more standard, purely hydrodynamical derivation.
Unitarity of the effective field theory  suggests an upper bound on the anomaly coefficient  in hydrodynamics.
\end{abstract}

%\pacs{\dots}% PACS, the Physics and Astronomy
\maketitle

%%%%%%%%%%%%%%%%%%%%%%%%%%%%%%%%%%%%%%%%%%%
%%%%%%%%%%%%%%%%%%%%%%%%%%%%%%%%%%%%%%%%%%%
\section{Introduction}
Relativistic hydrodynamics has recently attracted considerable attention from particle theorists. Part of the reason is that hydrodynamics turned out to be an adequate tool to describe
properties of the quark-gluon plasma produced at the RHIC and at the LHC. Additionally, AdS/CFT provided a rich theoretical laboratory for modeling hydrodynamics of strongly coupled field theories and calculating various transport coefficients.

One of the interesting recent results is that quantum anomalies---one of the most subtle phenomena in quantum field theory---may manifest in hydrodynamical transport coefficients.  Building  on earlier AdS/CFT results \cite{Erdmenger:2008rm,Banerjee:2008th} and field theory ones \cite{Vilenkin:1978hb,Vilenkin:1980fu,Fukushima:2008xe,Alekseev:1998ds}, ref.~\cite{Son:2009tf} presented a purely hydrodynamical derivation of this  effect. 

An interesting property of the anomalous transport is that it does not
lead to dissipation (in the form of entropy increase).
This suggests that it should
be relatively straightforward to incorporate anomalies into a
 field theoretical approach, which treats hydrodynamics as an
effective field theory for the Goldstone bosons associated with
spontaneously broken translations and global charges \cite{DGNR,ENRW}.
 If successful, this will be a non-trivial test for the field theory techniques.

Furthermore,  a complete understanding of the physical 
consequences of quantum anomalies in hydrodynamics is still lacking, and being worked out. 
A field theory description, if possible, may streamline this task.

The proposal of this paper is that anomalies should be reproducible by
adding Wess-Zumino (WZ)-like terms \cite{Wess:1971yu} to the fluid Lagrangian
of \cite{DGNR,ENRW}. The defining property of WZ terms is that they are invariant under the  symmetries only after integration by parts. In other words, even though the action is invariant in the presence of these terms, the Lagrangian density is not.

Here we report on our progress in implementing this proposal for
relativistic hydrodynamics in two
space-time dimensions. Two dimensions is curious: there is no
distinction between fluids and solids (see section \ref{sec:fluids}). But as we will see, it is
also particularly simple from the point of view of anomalies, 
for the relevant WZ term appears at the same derivative
level as the lowest-order, non-anomalous fluid Lagrangian. 

We start in section~\ref{sec:fluids} with reviewing the field theory 
description of non-anomalous hydrodynamics: we introduce the
symmetries defining the field theory, write down the action at the
leading order in derivatives, and establish the dictionary between
field theory and hydrodynamic observables. 
This is a brief review of what is presented in a companion paper
\cite{paper1}.

In section~\ref{WZsection} we present the Wess-Zumino term that can be added to this theory in accord with all its symmetries. In section~\ref{sec:gauging} we study the behavior of the resulting theory in the presence of an external gauge field and confirm that upon gauging, the Wess-Zumino term gives rise to the anomaly. In section~\ref{sec:dictionary} we establish the dictionary  between the field theory and hydrodynamics in the presence of the Wess-Zumino term. 

In section~\ref{sec:spectrum} we study the physical consequences of the Wess-Zumino term in the absence of an external gauge field. We find that that it gives rise to {\it half}  a new propagating degree of freedom, {\it i.e.} a purely left (right) moving mode---which we dub ``one-and-a-halfth sound". We calculate the propagation velocity of this mode and point out that unitarity of the field theory description sets an upper bound on the possible values of the anomaly coefficient.
In section \ref{sec:hydroderivation} we cross-check our field theory
results via a conventional hydrodynamical derivation of the
constitutive relations for anomalous hydrodynamics in two space-time
dimensions. We conclude in section~\ref{sec:last}.

\noindent
{\em Note added.}
While this manuscript was in preparation, two new preprints appeared which
also discuss fluid anomalies in two space-time dimensions
\cite{Kharzeev:2011ds,Loganayagam:2011mu}. Both papers derive the anomalous constitutive relations in
even space-time dimensions, at the level of hydrodynamical
equations, though not at the level of field theory and without studying the spectrum.

%%%%%%%%%%%%%%%%%%%%%%%%%%
%%%%%%%%%%%%%%%%%%%%%%%%%%
\section{Fluid Lagrangian  in two space-time dimensions}
\label{sec:fluids}
We start by reviewing the field theory description of non-anomalous fluids carrying conserved charges. We restrict to $(1+1)$ space-time dimensions. We only state the results here and refer the reader to \cite{paper1} for 
the derivation, which is  valid in any number of dimensions.
The dynamics of a perfect non-anomalous relativistic fluid in $(1+1)$ dimensions are described by a Lagrangian
of the form
\be \label{F(b,y)}
{\cal L}_0 = F(b,y) \; ,
\ee
 where 
\[
b\equiv \sqrt{\l \d_\mu\phi\r^2}\;,\qquad y \equiv {\epsilon^{\mu\nu}\d_\mu\psi\d_\nu \phi\over b}\;.
\]
We are using the $(-+)$-signature, and the sign convention for the  $\epsilon$-tensor is $\epsilon^{01}=1$. 
The scalar field $\phi$ describes a mapping of the physical space into the comoving fluid space.
This is a one-dimensional space, so, unlike in higher dimensions, there is no need to impose any symmetry acting on $\phi$ to ensure that the action describes a perfect fluid, rather than some general medium---a generic elastic medium 
in one spatial dimension ({\it i.e.}, in a tube or in a wire) can be thought of as a fluid.
Technically, this is because in one dimension the volume preserving diffs of \cite{paper1} are just constant shifts, $\phi \to \phi+a$.

On the other hand, on top of the $U(1)$ particle number symmetry
\be \label{U(1)}
\psi \to \psi + c \;,
\ee
which defines our conserved charge, one needs to impose the `chemical' shift symmetry \cite{sergey,paper1} 
\be \label{f(phi)}
\psi \to \psi + f(\phi) \; ,
\ee
to ensure that the action (\ref{F(b,y)}) does not depend on $(\d_\mu\psi)^2$. This dependence, if present, would indicate that we are actually dealing with a mixture of a normal fluid and a Bose condensate (superfluid component).

The shape of the function $F$ is determined by the fluid equation of state. Namely, the  dictionary establishing the correspondence between the field theory quantities and thermodynamical variables works in the following way.
We define the fluid two-velocity  as 
\be
\label{2velocity}
u^\mu={\epsilon^{\mu\nu}\d_\nu\phi\over b}
\ee
Then, from the field theory energy-momentum tensor 
\be
T_{\mu\nu} = (F_y y-F_b b) \, u_\mu u_\nu + (F - F_b b) \eta_{\mu\nu}
\ee
we read the expressions for the fluid density and pressure,
\be \label{rhoandp}
\rho = F_y y - F \; , \qquad p = F - F_b b \; .
\ee
Similarly, from the Noether current  associated with the shift symmetry acting on $\psi$,
\be \label{noether}
j^\mu = F_y u^\mu  ,
\ee
we deduce that the fluid charge density is 
 \be
\label{n=}
n = F_y \; .
\ee
Furthermore, we identify the identically conserved current
\be
\label{entropy_current}
J^\mu = b  \, u^\mu
\ee
with the 
entropy current, so that the entropy density is given by 
\be
\label{s=}
s = b \; .
\ee
Finally, by imposing the standard thermodynamic identities,
\be \label{thermoidentities} 
\rho + p = T s + \mu n \; , \qquad d \rho = T \, ds + \mu \, dn \; .
\ee
we find the expressions for the temperature and chemical potential of the fluid 
\be
\label{Tmu=}
T = -F_b \; , \qquad \mu = y \; , 
\ee
To summarize, the fluid Lagrangian is one of the most uncommon thermodynamic potentials,
\[
F(s,\mu)=n\mu-\rho
\]
which arises if one uses entropy density $s$ and chemical potential $\mu$ as the independent thermodynamic variables.

%%%%%%%%%%%%%%%%%%%%%%%%%%
%%%%%%%%%%%%%%%%%%%%%%%%%%
\section{Fluid Wess-Zumino Lagrangian}\label{WZsection}
It is straightforward to extend the above formalism to describe the fluid motion in the presence of an external gauge field $A_\mu$; for our purposes it is enough to
consider only the case of a fixed non-dynamical $A_\mu$. All that  one needs to do is to replace 
\be
\label{psigauging}
\d_\mu\psi\to \d_\mu\psi+A_\mu
\ee
everywhere in the fluid Lagrangian (\ref{F(b,y)}). General covariance of the fluid action in the presence of a non-trivial metric, implies that the fluid energy-momentum tensor (non)conservation in the presence of the external $A$ field
takes the form
\be
\label{dTA}
\d_\mu T_A^{\mu\nu}=F^{\nu\lambda}j_{A\lambda} \; .
\ee
in agreement with what one expects in hydrodynamics. The subscript $A$
here is a reminder that the energy-momentum tensor and the current in
(\ref{dTA}) are obtained from the energy-momentum tensor and the
Noether current of the fluid in the absence of the external field,
followed by applying the replacement (\ref{psigauging}). 

As was realized recently \cite{Son:2009tf}, hydrodynamics gets modified in an interesting and controllable way when the particle number symmetry \eqref{U(1)} is anomalous, meaning that  the corresponding current is 
no longer conserved in the presence of an external gauge field. In a space-time with an even dimensionality $D$, the current's
divergence is proportional to the wedged product $F\wedge\dots\wedge F$ of $D/2$ field-strength two-forms.  
 
 Interestingly, even though the effects of anomalies in $D= 4$ (the case considered in  \cite{Son:2009tf}) arise at the same order in the derivative expansion as various dissipative phenomena (shear viscosity, bulk viscosity, and
 conduction), they do not lead to an entropy growth on their own. The same holds for higher dimensions as well \cite{Kharzeev:2011ds,Loganayagam:2011mu}.
 This suggests that it should be possible to incorporate them in our Lagrangian formalism without introducing extra degrees of freedom, which would be needed to describe dissipation. 
 
 Furthermore, the presence of an anomaly in   \cite{Son:2009tf} has  physical consequences even in the absence of an external field. For instance, it  gives rise to a contribution to the particle current proportional to the fluid vorticity.
 This is reminiscent of what happens in the pion chiral Lagrangian, where in order to reproduce the anomalies arising as a result of gauging the flavor symmetries one introduces a Wess-Zumino term in the pion Lagrangian, which gives rise to physical processes, such as $\pi\pi\to KKK$ scattering even in the absence of gauge fields.
 
 The key property of the Wess-Zumino term for reproducing the
 anomalies is that it is invariant under the flavor symmetries only at
 the level of the action, {\it i.e.} the Lagrangian itself acquires a
 variation which is a total derivative. As a result the invariance may
 get lost upon gauging, implying the anomaly. 

To see whether the same mechanism works in hydrodynamics we need to find an additional term in the fluid Lagrangian, which is non-invariant  under some of the fluid symmetries, but gives rise to a variation which is a total derivative. In this paper we will accomplish this in the simplest case of $D=(1+1)$ hydrodynamics, where the only non-trivial  symmetry is the chemical symmetry (\ref{f(phi)}).

Another simplification occurring  in the $D=2$ case is that the anomalous divergence is proportional simply to $\epsilon^{\mu\nu}F_{\mu\nu}$. Unlike in higher dimensions, where
anomalies arise at higher orders in the derivative expansion, this term is of the same order as the leading order terms associate with (\ref{F(b,y)}). So, to construct a Wess-Zumino term, we need to find an expression which {\em (i)} changes under the chemical symmetry (\ref{f(phi)}) by a total derivative, and {\em (ii) }involves exactly one derivative per each $\psi$ and $\phi$ field.

Without further ado let us present the action  with the required properties:
\be
\label{WZaction}
S_{\rm WZ}=-C\int d^2x\; u^\mu\d_\mu\psi \, \tilde{u}^\nu\d_\nu\psi\;,
\ee
where $u^\mu$ is the fluid two-velocity,
and
\[
\tilde{u}^\mu \equiv \epsilon^{\mu\nu}u_\nu=b^{-1}\d^\mu\phi 
\]
is orthogonal to it, and normalized to one.
This action is not of the form (\ref{F(b,y)}), so there is no invariance under (\ref{f(phi)}) at the level of the Lagrangian. However, it is straightforward to check that the corresponding variation is a total derivative,
\begin{align}
\delta S_{\rm WZ} & =-C\int d^2x\;u^\mu\d_\mu\psi \, \tilde{u}^\nu\d_\nu f(\phi) \nonumber  \\
& = -C\int d^2x\; \epsilon^{\mu\nu}\d_\mu\psi \, \d_\nu f(\phi)           
\label{invcheck}
\end{align}
where we made use of the identity
\be
\label{epsidentity}
u^\mu\tilde{u}^\nu=\epsilon^{\mu\nu} + \tilde{u}^\mu u^\nu\;.
\ee
Consequently, the action
\be
\label{fullWZaction}
S=S_0+S_{\rm WZ}
\ee
satisfies all the symmetries expected for the perfect fluid
action. One expects that there should exist an extension of the dictionary outlined in Sec.~\ref{sec:fluids} between our field theory and (anomalous) hydrodynamics
applicable in the presence of the Wess-Zumino term (\ref{WZaction}). To establish this dictionary, it is instructive to consider first what happens with the system (\ref{fullWZaction}) in the presence of an external gauge field.
Along the way we will also check that the Wess-Zumino term (\ref{WZaction}) indeed gives rise to the anomaly.

%%%%%%%%%%%%%%%%%%%%%%%%%%
%%%%%%%%%%%%%%%%%%%%%%%%%%
\section{Gauging the  Wess-Zumino term}
\label{sec:gauging}
Let us try first to introduce the external field following the rule (\ref{psigauging}).  
This  ``naive"  gauging leads to the action
\be
S_{\rm ng}[\phi,\psi,A_\mu] \equiv S[\phi,\d_\mu\psi+A_\mu]\;.
\ee
The problem with this gauging is that in the presence of the Wess-Zumino term (\ref{WZaction}), it does not preserve the chemical symmetry (\ref{f(phi)}).
Indeed, following the same steps as in (\ref{invcheck}) we obtain  a non-trivial variation of the form
\[
\delta S_{\rm ng}
=-C\int  \epsilon^{\mu\nu}A_\mu\d_\nu f(\phi)\;.
\]
The invariance under the chemical symmetry can be restored by sacrificing gauge invariance, and postulating that the correct action
describing the fluid in an external $A$ field  is
\be
\label{Sg}
S_{\rm g}=S_{\rm ng}+C\int  \epsilon^{\mu\nu}A_\mu\d_\nu\psi\;.
\ee
This action is no longer invariant under the gauge symmetry:
its anomalous variation is
\[
\delta S_{\rm an}=\sfrac12 C\int  \alpha \epsilon^{\mu\nu}F_{\mu\nu}\;,
\]
where $\alpha$ is the gauge transformation parameter, $A_\mu\to A_\mu+\d_\mu\alpha$, $\psi\to\psi-\alpha$.
Notice that this anomalous variation is independent of the dynamical `matter fields' of the theory---it only depends on the external gauge field. This allows one to apply the conventional argument for
why gauging an anomalous symmetry is a sensible and well-defined procedure. Namely, one can introduce an additional ``spectator" sector, which is not directly coupled to our fluid, whose only purpose in life is to cancel the anomaly in the full theory.

Let us see now how the presence of the anomaly affects the fluid equations.
First, note that the extra term in \eqref{Sg} does not contribute to the energy-momentum tensor.
However, this term does contribute to the energy-momentum (non)conservation. Namely, the invariance of the gauged action under a combined
diff transformation of the dynamical fields, the metric and the external gauge field implies,
%\be
%\label{diffS}
%\int -{\delta S_g\over \delta g^{\mu\nu}}\l\d^\mu\xi^\nu+\d^\nu\xi^\nu\r+{\delta S_g\over \delta A_\mu}\l\d_\nu A_\mu\xi^\nu+\d_\mu\xi^\nu A_\nu\r+\dots\;,
%\ee
%where ellipses stand for terms involving variations of dynamical fields under diffs. These terms vanish on-shell, so that integrating (\ref{diffS}) by parts we arrive at
\be
\d_\mu T^{\mu}  {}_{\nu}={\delta S_g\over \delta A_\mu}F_{\nu\mu}-A_\nu\d_\mu{\delta S_g\over\delta A_\mu}\;.
\ee
Note that the gauged action (\ref{Sg}) is invariant under shifts of $\psi$; the corresponding conserved Noether current is equal to
\be
\label{Noether}
j^\mu_N={\delta S_g\over\delta A_\mu}-C\epsilon^{\mu\nu}A_\nu-C\epsilon^{\mu\nu}\d_\nu\psi\;.
\ee
Conservation of this current implies,
\be
\d_\mu{\delta S_g\over\delta A_\mu}=\sfrac12 {C} \epsilon^{\mu\nu}F_{\mu\nu}
\ee
so that the energy-momentum conservation takes the  form,
\be
\label{dTMN}
\d_\mu T^{\mu\nu}=F^{\nu\mu}{\delta S_g\over\delta A^\mu}- \sfrac12 {C} A^\nu\epsilon^{\mu\lambda}F_{\mu\lambda}\;.
\ee
By making use of the identity 
\[
\sfrac12 A^\nu\epsilon^{\mu\lambda}F_{\mu\lambda}=- F^{\nu\alpha}\epsilon_{\alpha\beta}A^\beta
\]
the energy-momentum (non)conservation can be presented in the traditional hydrodynamical form
\be
\d_\mu T^{\mu\nu} = F^{\nu\mu}j_\mu
\ee
where the hydrodynamical current is
\be
\label{jhydro}
j^{\mu}={\delta S_g\over\delta A^\mu}+C\epsilon_{\mu\nu}A^{\nu}=F_y u^\mu-2C\, y \, \tilde{u}^\mu\; ,
\ee
which is manifestly gauge invariant ($F$ and $y$ here and henceforth stand for their gauged versions $F_A$ and $y_A$, according to the notation of sect.~\ref{WZsection}).
As a consequence of the anomaly the gauge-invariant hydrodynamical current $j^\mu$ is different from the conserved, but gauge non-invariant Noether current (\ref{Noether}).
The anomalous divergence of the hydrodynamical current is equal to
\be
\d_\mu j^{\mu}=C\epsilon^{\mu\nu}F_{\mu\nu}\;,
\ee
as desired.

%%%%%%%%%%%%%%%%%%%%%%%%%%
%%%%%%%%%%%%%%%%%%%%%%%%%%
\section{Hydrodynamic dictionary}
\label{sec:dictionary}
It is straightforward now to establish the dictionary between our field theory with the Wess-Zumino term and the conventional description of (anomalous) hydrodynamics.
As before, we identify the normalized velocity of comoving volume elements with the two-velocity of the fluid, see Eq.~(\ref{2velocity}).
Applying the same logic as in the non-anomalous case \cite{paper1} we identify the identically conserved current
\[
J^\mu=\epsilon^{\mu\nu}\d_\nu\phi
\]
with the entropy current, which gives us 
\[
s=b
\]
for the entropy density. 
Note that for anomalous hydrodynamics, the  rest frames for mass, charge, and entropy, are in general all different from each other.
We see that in the field theory description the ``entropy frame"---which identifies the fluid velocity with the direction of the entropy flow---is the most natural one.
The charge density $n$ is defined as the projection of the hydrodynamical current (\ref{jhydro}) onto the fluid velocity,
\be
n=-j_\mu u^\mu=F_y 
\ee
---unmodified w.r.t~eq.~\eqref{n=}.
To identify the field theory operators corresponding to the fluid density and pressure, we need the contribution of the Wess-Zumino term to the energy-momentum tensor,
\be
\label{TmunuWZ}
T^{\rm WZ}_{\mu\nu}=-Cy^2(\tilde{u}_\mu u_\nu+\tilde{u}_\nu u_\mu)\;.
\ee
We see that the contractions of the  Wess-Zumino energy-momentum with both $u^\mu u^\nu$ and $\eta^{\mu\nu}$ are zero. Consequently, there is no anomaly contribution to the fluid density and pressure either. 
Note also that $T^{\rm WZ}_{\mu\nu}$ is traceless by virtue of the Weyl
invariance of the Wess-Zumino term.
Then, by applying the thermodynamic identities
\be \label{thermoidentities1} 
\rho + p = T s + \mu n \; , \qquad d \rho = T \, ds + \mu \, dn \; ,
\ee
we find that the expressions for the fluid temperature and chemical potential also remain the same, eq.~(\ref{Tmu=}). Consequently, the whole dictionary establishing the relation between the hydrodynamic variables and the field theory ones remains unchanged in the presence of the Wess-Zumino term.
The only consequence of this term is the change of the constitutive relations for the fluid current and energy-momentum. When written in  fluid variables, the new anomalous contributions to these quantities take the form
\begin{align}
\label{fieldtheoryDT}
& \Delta T^{\mu\nu} = -C \mu^2  (u^\mu \tilde u^\nu + u^\nu \tilde
u^\mu) \\
& \Delta j^\mu = - 2C \mu  \tilde u^\mu \; . \label{fieldtheoryDj}
\end{align}
In section \ref{sec:hydroderivation} we will confirm this result by a direct hydrodynamical calculation.

A microphysical interpretation of the chiral vortical effect of ref.~\cite{Son:2009tf} is still lacking. On the other hand, the microscopic origin of our correction to the current---eq.~\eqref{fieldtheoryDj}---is quite clear. In $1+1$ dimensions, there is no spin, and chirality for fermions takes a degenerate form: left-handedness is the same as left-movingness. As a result, at final chemical potential for a chiral charge, there are more left-moving fermions than right-moving ones (or viceversa). This mismatch leads to a net current for this charge. The effect is proportional to the anomaly coefficient---which just counts the number of fermion species involved democratically in the charge under consideration. The appearance of the chemical potential may seem more puzzling at first: given the physical argument above, one would expect the effect to be proportional to the charge density $n$. However the average speed at which the net charge moves is determined by thermodynamical equilibrium---one certainly cannot model the effect by considering free-streaming left-moving and right-moving particles.

%%%%%%%%%%%%%%%%%%%%%%%%%%
%%%%%%%%%%%%%%%%%%%%%%%%%%
\section{Spectrum of perturbations}
\label{sec:spectrum}
To see some of the physical consequences of the fluid Wess-Zumino term, let us study the spectrum of linear excitations around a static homogeneous fluid ground state. We will restrict ourselves to the case of zero external field.
Then the perturbed fluid configuration can be written as 
\[
\phi=s \cdot (x+\pi) \;,\qquad\psi=\mu \cdot (t+\chi)\;,
\]
where $s$ and $\mu$ are the unperturbed entropy density and chemical potential.
The quadratic Lagrangian for small perturbations $\pi$, $\chi$ reads
\begin{align}
{\cal L}_2 & =\sfrac12 (\mu F_y-sF_b)\dot{\pi}^2+\sfrac12 s^2F_{bb}\pi'^2+\sfrac12 \mu^2F_{yy}\dot{\chi}^2 \nonumber \\
& + (2s\mu F_{yb}+\mu F_y)\dot{\chi}\pi' \nonumber \\
& -C\mu^2(\dot{\pi}\pi'+\dot{\chi}\chi')+2C\mu^2\dot{\pi}\dot{\chi} \; ,
\label{L2}
\end{align}
where the derivatives of $F$ are all computed on the background configuration.
This is a coupled second-order system of two fields, so in general one expects to find four propagating modes for each value of  the spatial momentum $k$---two left-movers and two-right movers. 
Without our Wess-Zumino term, the fluid Lagrangian is invariant under the spatial parity $x\to-x$, $\phi\to-\phi$,
which implies that left- and right-movers have the same propagation velocity.

Furthermore, 
the chemical symmetry implies that there should be a mode with vanishing propagation velocity, {\it i.e.}, with a dispersion relation $\omega=0$.
Indeed, $\chi=\chi(x)$ is a solution of (\ref{L2}). 
As a result, without the Wess-Zumino term one is left with a single left-moving and a single  right-moving phonon with equal propagation velocities.
If one were to include dissipative effects, the non-propagating $\omega=0$ modes would acquire an imaginary contribution to their dispersion relation and would describe charge diffusion.

A new phenomenon arising in the presence of the Wess-Zumino term, is that {\em one} of these `frozen' modes starts propagating.
This is still compatible with the chemical symmetry: $\chi = \chi(x)$ is still a solution. Only, since the  Wess-Zumino term breaks spatial parity, left- and right-movers may have different 
propagation velocity.  
 A direct inspection of the dispersion relation following from (\ref{L2}) confirms the presence of one $\omega=0$ mode and three propagating modes.
The expressions for their propagation velocities are somewhat cumbersome and non-illuminating in general, so we only discuss here  two limiting cases. 
 
For small $C$, one finds a pair of left- and right-moving phonons with approximately equal propagation speeds,
 \[
 \omega_{L,R}= v_{s \, L, R} \, k
 \]
where $k$ is the absolute value of the momentum and
\[
v_{s \, L, R}^2={(2s F_{by}+F_y)^2-s^2F_{yy}F_{bb}\over F_{yy}(F_ys- \mu F_b)} \pm {\cal{O}}(C) \; .
\]
These are the ordinary sound waves, slightly corrected by the presence of the anomaly.
The ${\cal{O}}(C)$ corrections to the speed of sound are different (opposite, in fact) for left- and right-moving phonons.

One top of these, one finds a   mode that propagates only in one direction---one-and-a-halfth sound. This mode is similar to the chiral magnetic wave in four dimensions \cite{Kharzeev:2010gd}, but in two dimensions it appears even in the absence of an external field.
 It has a linear dispersion relation
\[
\omega = v_{\rm WZ} \,k \; ,
\]
with propagation speed equal to 
\[
v_{\rm WZ}=2C {s^2 F_{bb}\over s^2 F_{yy}F_{bb}-(2s F_{by}+F_y)^2}+{\cal{O}}(C^2) \; .
\]

Another limit which is straightforward to look at, is when the action is dominated by the Wess-Zumino term, $F \to 0$. In such a case the quadratic Lagrangian \eqref{L2} reduces simply to 
\[
{\cal L}_2 \to \sfrac12\mu^2C\l\dot\pi_+^2-\dot\pi_-^2-\dot\pi_+\pi_+'-\dot\pi_-\pi_-'\r \; ,
\]
where $\pi_\pm=\pi\pm\chi$. Each of the fields $\pi_\pm$ describes one mode with an $\omega=0$ dispersion relation, and one  propagating at the speed of light. However, the theory is sick in this case, as one of the fields $\pi_\pm$ is a ghost---it has a wrong-sign kinetic term.
This indicates that there is an
upper bound on the anomaly coefficient $C$ following from banning ghosts.
For instance,  requiring that the quadratic action (\ref{L2}) is ghost-free at zero momentum implies
 \be
 \label{dndscondition}
 4C^2<(\mu F_y - s F_b)F_{yy}/\mu^2 ={\rho+p \over \mu^2}\l {\d n\over \d\mu}\r_s
 \; .
 \ee
  
Banning other pathologies---such as superluminal modes and classical instabilities---may lead to more constraints (cf.~the analysis of \cite{Dubovsky:2004sg,DGNR}). For instance,
another necessary condition for the absence of  pathologies follows from requiring that the energy-momentum tensor of the anomalous fluid satisfies the null energy condition, {\it i.e.},
 \be
 \label{NEC}
 T_{\mu\nu}n^\mu n^\nu \ge 0
 \ee
 for any null vector $n^\mu$. The results of  \cite{DGNR} are directly applicable to our case and imply that a violation of (\ref{NEC}) is necessarily accompanied
 by ghosts or classical instabilities or superluminal modes. The two independent null vectors in two dimensions are $u^\mu\pm\tilde u^\mu$, so that in the presence 
 of the Wess-Zumino contribution (\ref{TmunuWZ}) the null energy condition reads 
 \be
 \label{NECcondition}
 |C|\le{\rho+p\over 2\mu^2} \; ,
 \ee
which in general is neither stronger nor weaker than \eqref{dndscondition}---just different.
A comprehensive analysis of  stability and sub-luminality may provide  stronger constraints than the necessary conditions (\ref{dndscondition}),  (\ref{NECcondition}).

This example shows one advantage of having the field theory description: the presence of a ghost is not obviously detectable in the hydrodynamical language, and one needs a field theory to deduce that the coefficient $C$ cannot be arbitrarily large.

%%%%%%%%%%%%%%%%%%%%%%%%%%
%%%%%%%%%%%%%%%%%%%%%%%%%%
\section{Hydrodynamic Derivation}
\label{sec:hydroderivation}
%Notes 2011 II p. 68 - 84.
We follow \cite{Son:2009tf} by giving a hydrodynamic derivation
in the momentum frame,
{\it i.e.}~where $\Delta T^{\mu\nu} = 0$. 
At the end, we rotate back to the more natural frame from the field theory
point of view, {\em i.e.}~the entropy frame where $\Delta s^\mu = 0$.
The energy-momentum, charge current, and entropy current are
\begin{align}
T^{\mu\nu} & = (\rho_0 + p_0) u^\mu_0 u^\nu_0 + p_0 \, \eta^{\mu\nu} \\
j^\mu & = n_0 u^\mu_0 + \xi_j \tilde u^\mu_0 \\
s^\mu & = s_0 u^\mu_0 + \xi_s \tilde u^\mu_0 \, ,
\end{align}
where we use the subscript $0$ to denote momentum-frame quantities.
Here, $\Delta j^\mu = \xi_j \tilde u^\mu_0$ and
$\Delta s^\mu = \xi_s \tilde u^\mu_0$ are the anomaly
corrections to the charge and entropy currents.
The equations of motion are
\begin{eqnarray}
\partial_\mu T^{\mu\nu} = F^{\nu\alpha} j_\alpha , \quad
\partial_\mu j^\mu = C \epsilon^{\mu\nu} F_{\mu\nu}  , \quad
\partial_\mu s^\mu = 0 \, .
\end{eqnarray}
In particular, entropy is conserved because the anomaly should not introduce any dissipation, which  we are therefore neglecting altogether.

The external electric field $E$ is related to $F_{\mu\nu}$ by
$F_{\mu\nu} = E \epsilon_{\mu\nu}$, so that
$F^{\nu\alpha} j_\alpha = \xi_j E u^\nu_0 + n E \tilde u^\nu_0$
and $\epsilon^{\mu\nu} F_{\mu\nu} = - 2 E$.
Note the useful relations:
$u_0^2 = -1$, $\tilde u_0^2 = 1$, $u^\mu_0 \tilde u^\nu_0 - u^\nu_0 \tilde u^\mu_0
= \epsilon^{\mu\nu}$, and $\tilde u^\mu_0 \tilde u^\nu_0 - u^\mu_0 u^\nu_0 = \eta^{\mu\nu}$.

Projecting the energy-momentum equation along the fluid flow,
and using the thermodynamic relations $\rho_0 + p_0 = T_0 s_0 + \mu_0 n_0$
and $d\rho_0 = T_0 ds_0 + \mu_0 dn_0$, we obtain:
\begin{eqnarray}
- T_0 \partial \cdot (s_0 u_0) - \mu_0 \partial \cdot (n_0 u_0) = - \xi_j E \, ,
\end{eqnarray}
which implies
\begin{eqnarray}
\label{e1}
T_0 \partial \cdot (\xi_s \tilde u_0) + \mu_0 \partial \cdot (\xi_j \tilde u_0) + 2 \mu_0 C E
= - \xi_j E \, ,
\end{eqnarray}
upon using the charge and entropy equations.

Projecting the energy-momentum equation orthogonally to the flow, we have
\begin{eqnarray}
\label{e2}
(\rho_0 + p_0) \partial \cdot \tilde u_0 + \tilde u_0 \cdot \partial
p_0 = n_0 E \, .
\end{eqnarray}
This can be used to remove $\partial \cdot \tilde u_0$ from
Eq. (\ref{e1}), giving
\begin{align}
\tilde u_0 \cdot &  
\left( T_0 \partial \xi_s + \mu_0 \partial \xi_j
- {T_0 \xi_s + \mu_0 \xi_j \over \rho_0 + p_0} \partial p_0 \right) \nonumber \\ 
&= -E \left(2 \mu_0 C + \xi_j + {n_0 (T_0 \xi_s + \mu_0 \xi_j) \over \rho_0 + p_0} \right)
\end{align}
Since $E$ and $\tilde u_0$ are arbitrary and independent, we demand each side to vanish independently.
Thus,
\begin{eqnarray}
\label{Eeqt}
2\mu_0 C + \xi_j + {n_0 T_0 \over \rho_0 + p_0} 
\left(\xi_s + {\mu_0 \over T_0} \xi_j \right) = 0 \, ,
\end{eqnarray}
and
\begin{eqnarray}
\label{tildeueqt}
\partial \left(\xi_s + {\mu_0 \over T_0} \xi_j \right)
- \xi_j \partial {\mu_0 \over T_0} = \left( \xi_s + {\mu _0\over T_0} \xi_j \right)
 {\partial p_0 \over \rho_0 + p_0}
\, .
\end{eqnarray}
Comparing Eq. (\ref{tildeueqt}) with the thermodynamic relation
$d T_0 + n_0 T_0^2 /(\rho_0 + p_0) \,d(\mu_0/T_0) = T_0 \, d p_0/(\rho_0 + p_0)$,
one can see that
\be
\xi_s + \mu_0 \xi_j / T_0 = T_0 g(\mu_0/T_0)
\ee
---this guarantees their mutual
consistency at constant $\mu_0/T_0$,
while consistency at constant $p_0$ means the function $g$ obeys
\begin{eqnarray}
{n_0 T_0^2 \over \rho_0 + p_0} = {T g' - \xi_j \over g} \; .
\end{eqnarray}
Substituting $\xi_j$ from Eq.~(\ref{Eeqt}), we thus obtain $g' = - 2 C \mu_0/T_0$,
which implies $g = - C (\mu_0/T_0)^2 + d$, where $d$ is an integration constant
(its analog in $3+1$ was pointed out by \cite{neimanoz}). We therefore arrive at the result:
\begin{eqnarray}
\label{momframexij}
\xi_j = \left( {n_0 \mu_0^2 \over \rho_0 + P_0} - 2 \mu_0 \right) C
- {n_0 T_0^2 \over \rho_0 + p_0} d\, ,
\end{eqnarray}
\begin{eqnarray}
\label{momframexis}
\xi_s =  {\mu_0^2 s_0 \over \rho_0 + p_0}  C
+ T_0 \left( 1 + {n_0 \mu_0 \over \rho_0 + p_0} \right) d
\end{eqnarray}

This result is expressed more simply in the entropy frame.
Suppose for the moment that $\xi_j$ and $\xi_s$ are small, so that
$s^\mu = s_0 u^\mu_0 + \xi_s \tilde u_0^\mu = s u^\mu$ means
$s \sim s_0$ and $u^\mu \sim u^\mu_0 + \xi_s \tilde u_0^\mu / s_0$,
i.e. enforcing $u^2 = -1$ would only introduce terms of $O(\xi_s^2)$.
Here, we use symbols without the subscript $0$ to denote entropy frame quantities.
It is straightforward to see that in the frame where $s^\mu =
s u^\mu$:
\begin{align}
T^{\mu\nu} & =  (\rho + p) u^\mu u^\nu + p \eta^{\mu\nu} \nonumber \\
& - \left( \mu^2 C + (T^2 + 2nT\mu/s) d\right) (u^\mu \tilde u^\nu
+ u^\nu \tilde u^\mu) \\
j^\mu & = n u^\mu - 2 \left( \mu C + {nT \over s} d \right) 
\tilde u^\mu \, .
\end{align}
This entropy frame result looks strikingly simple especially if one ignores the
integration constant, {\em i.e.}~sets $d = 0$.
In fact, the rotation from momentum frame to entropy frame can be done
non-perturbatively. The exact mapping between quantities in
the two frames, {\it in the absence of integration constant terms}, is
\begin{align}
 & u^\mu  =  {\bar c} \, u^\mu_0 + {\bar s} \, \tilde u^\mu_0 \; , \quad
\tilde u^\mu = {\bar s} \,  u^\mu_0 + {\bar c} \,  \tilde u^\mu_0 \nonumber \\
&s  = s_0/\bar c \; , \quad \mu = \bar c \, \mu_0 \; , \quad
T = \bar c \, T_0 \; , \quad p = \bar c^2 p_0 + \bar s^2 \rho_0 \quad
\nonumber \\
& \rho = \bar s^2 p_0 + \bar c^2 \rho_0 \; , \quad
n = {1\over \bar c}\left( n_0 + 2 \bar s^2 {\rho_0 + p_0 \over
    \mu_0}\right) 
\end{align}
where the boost parameters $\bar c \equiv \cosh \eta$ and $\bar s \equiv \sinh \eta$ satisfy:
\begin{eqnarray}
\bar c^2 - \bar s^2 = 1 \; , \qquad {\bar s \over \bar c} = {\mu_0^2 \over \rho_0 + p_0} C \, .
\end{eqnarray}
It can be checked that the thermodynamic relations work out correctly
for the entropy frame quantities
\footnote{This is non-trivial. For instance, if $d$ were non-zero,
using the more general expressions for $n =
\bar c n_0 - \bar s \xi_j = (n_0 - \bar s \big(\bar c \xi_j - \bar s
n_0)\big)/\bar c$, and $\bar s /\bar c = \xi_s / s_0$, the above mapping
would not satisfy $\rho + p = Ts + \mu n$.
},
and that the {\it exact} anomaly
corrections in entropy frame are
$\Delta T^{\mu\nu} = - \bar c \bar s (\rho_0 + p_0) (u^\mu \tilde
u^\nu + u^\nu \tilde u^\mu)$ and
$\Delta j^\mu = (\bar c \xi_j - \bar s n_0) \tilde u^\mu$, giving:
\begin{eqnarray}
\Delta T^{\mu\nu} = - \mu^2 C (u^\mu \tilde u^\nu + u^\nu \tilde
u^\mu)
\; , \quad \Delta j^\mu = - 2 \mu C \tilde u^\mu \, .
\end{eqnarray}
This agrees nicely with the field theory result Eqs.~\eqref{fieldtheoryDT}, \eqref{fieldtheoryDj}.

What is the interpretation of the integration constant terms, those
proportional
to $d$ in Eqs. (\ref{momframexij}) \& (\ref{momframexis})?
One can get a hint by rotating to the charge frame, where $j^\mu = n
u^\mu$, instead of the entropy frame. 
{\it In the absence of
anomaly}, i.e.~setting $C=0$, one
obtains:
\begin{eqnarray}
\label{integconstDTDs}
%\Delta T^{\mu\nu} = [(2\mu [\rho + P]/n - \mu^2) C + T^2 d]
%(u^\mu \tilde u^\nu + u^\nu \tilde u^\mu) \quad , \quad 
%\Delta s^\mu = 2 (\mu s C/n + T d) \tilde u^\mu \, .
\Delta T^{\mu\nu} = T^2 d \,
(u^\mu \tilde u^\nu + u^\nu \tilde u^\mu) \; , \quad 
\Delta s^\mu = 2  T d \, \tilde u^\mu \, ,
\end{eqnarray}
where we use symbols with no subscript $0$ to denote charge frame quantities.
Thus, the integration constant terms appear particularly simple in
this frame. A field theory realization is:
\begin{eqnarray}
S = \int d^2 x \, F(b,y) + d (u \cdot \partial \psi) (\tilde u
\cdot \partial \psi) + \epsilon^{\mu\nu} \partial_\nu \phi A_\nu \, ,
\end{eqnarray}
with the modified dictionary: $s^\mu = F_y u^\mu + 2d \, y \, \tilde u^\mu$, $s =
F_y$, $j^\mu = \epsilon^{\mu\nu} \partial_\nu \phi = b u^\mu$, $n =
b$, $T = y = u \cdot \partial \psi$, $\mu = - F_b$. 
It can be verified that this dictionary reproduces
Eqs. (\ref{integconstDTDs}), and gives $\partial_\mu T^{\mu\nu} =
F^{\nu\alpha} j_\alpha$ and $\partial_\mu j^\mu = 0$. 
Consequently, when the current is non-anomalous, the integration constant exists due to the possibility of establishing an alternative dictionary.
This possibility is related to some flexibility in how to identify  the entropy current (although, arguments presented in \cite{paper1} favor our canonical choice (\ref{entropy_current})). This flexibility is absent in the presence of the anomaly, when  (\ref{entropy_current}) is the only current which is conserved in the presence of an external field.

It was argued in \cite{bhatt} that one of the two integration constant terms in
$3+1$ should vanish because it violates CPT invariance, while the
other is allowed.
It is clear our hydrodynamic result for the integration constant
in $1+1$,  Eq. (\ref{integconstDTDs}), is consistent with CPT 
(under which $u^\mu \rightarrow -u^\mu$, $\tilde u^\mu \rightarrow
-\tilde u^\mu$, $n \rightarrow -n$, $\mu \rightarrow -\mu$, 
while other relevant quantities remain unchanged such as $j^\mu$ and $T^{\mu\nu}$),
and indeed our field theory realization confirms it.
It is not obvious what field theory $+$ dictionary would reproduce
both the anomaly and the integration constant at the same time.
\section{Concluding remarks}
\label{sec:last}
To summarize, our results demonstrate that in two space-time dimensions it is possible to incorporate the effect of anomalies in hydrodynamics
by introducing a Wess-Zumino term in the effective field theory description of fluids. It is straightforward to extend these results to an arbitrary number of global charges.

It would be interesting to establish a relation between the existence of Wess-Zumino terms and the topology of  field space in the fluid effective field theory.
For a pion chiral $\sigma$-model this allows, in particular, to prove the quantization of the coefficient in front of the Wess-Zumino term \cite{Witten:1983tw}. At the current level of understanding we cannot prove this for a fluid Wess-Zumino term. However, the field theory description already strongly suggests that this coefficient cannot get renormalized. Indeed, one can apply the following argument commonly used to prove various non-renormalization theorems in field theory. If this coefficient  depended on any coupling constants, we could deform the theory by allowing these couplings to vary slightly in space-time. However, this would ruin the chemical symmetry, which for   the   Wess-Zumino term
holds only after integration by parts.

Given how simply and naturally things work in the two-dimensional case, one may expect that it should be straightforward to generalize our results to anomalies in higher dimensions.
Our investigations so far suggest that the full story is a bit  subtler
than for the two-dimensional case.
Let us nevertheless present here what we expect to be very close to the correct answer.

It is convenient to use the language of differential forms and to decompose the gradient of $\psi$ as
\[
d\psi=(d \psi)_u+\alpha
\]  
where $(d\psi)_u\equiv -y \, u_\mu dx^\mu$ is the projection of $d\psi$ on the direction of the fluid flow. This projection is invariant under the chemical symmetry, while the one-form $\alpha$ transforms as a gauge potential, 
$\alpha\to\alpha+ df(\phi)$. Then the Wess-Zumino term (\ref{WZaction}) can be written as
\[
S_{\rm WZ}=C\int (d\psi)_u\wedge d\psi=C\int (d\psi)_u\wedge \alpha=C\int d\psi\wedge\alpha
\]
The very last way of writing the Wess-Zumino term makes its invariance under the chemical symmetry manifest---the variation of the first factor gives rise to $df(\phi)\wedge\alpha$, which is zero because both one-forms here
are proportional to $d\phi$ (recall that $\alpha$ is orthogonal to the flow), while the variation of $\alpha$ gives rise to a total derivative, $d\psi\wedge d f(\phi)$. A natural generalization of this expression to $D=2n+2$ dimensions appears to be 
\be
\label{S2np2}
S_{2n+2}=\int d\psi\wedge\alpha\wedge (d\alpha)^n\;.
\ee
In particular, in $D=3+1$ this would give 
$
\int  y^2\epsilon^{\mu\nu\lambda\rho} \, \d_\mu\psi \, u_\nu\d_\lambda u_\rho
$. Unfortunately, these expressions are not invariant under the chemical symmetry, because the invariant two-form $d\alpha$ in general has non-vanishing non-comoving ($d\phi\wedge u_\mu dx^\mu$) components, so that the variation of $d\psi$ results in a non-vanishing variation of (\ref{S2np2}). We are not certain at the moment whether this indicates that one should give up the full chemical symmetry in higher dimensions, or that a more sophisticated Wess-Zumino term is required.

Interestingly, the fluid Lagrangian allows a Wess-Zumino term in any {\em odd} number of dimensions, which {\it is} invariant under the chemical symmetry,
\be
S_{2n+1}=\int \alpha\wedge (d\alpha)^n\;.
\ee 
This term does not lead to any anomalies upon gauging, as expected in odd number of dimensions, but it would be interesting to study its other physical consequences. 
We plan to address this as well as the fate of the fluid Wess-Zumino terms in
even $D>2$ dimensions in the near future. 

Finally, it would  be very interesting to see if the anomaly discussed in
this paper and its physical manifestations such as the 1.5th
sound, are realized in one-(spatial)-dimensional condensed matter
systems 
with massless fermionic excitations, e.g. \cite{Alekseev:1998ds}.

\noindent
{\em Acknowledgements.}
We would like to thank Dam Thanh Son for very useful discussions,
and for posing the question of
formulating a field theory description of fluids with anomalies. We also thank
Andrei Gruzinov, Rafael Porto, Slava Rychkov and Matias Zaldarriaga
for useful discussions.
We are supported in part by the DOE under contracts DE-FG02-92-ER40699 (LH, AN)
and DE-FG02-11ER1141743 (AN), and by NASA under contract NNX10AH14G (LH, AN).
LH thanks HKU and the IAS at HKUST for hospitality. AN thanks the Laboratoire de Physique Th\'eorique at ENS
for hospitality.

\end{document}